\PassOptionsToPackage{dvipsnames,table,svgname,hyperref}{xcolor}
\documentclass[sigconf]{acmart}
\usepackage{amsmath}
\usepackage{multirow}
\usepackage{bm}
\usepackage{makecell}
\usepackage{xspace}
\usepackage{soul}
\usepackage{cleveref}
\usepackage{balance}

\definecolor{green}{rgb}{0.13, 0.55, 0.13}
\definecolor{red}{rgb}{0.8, 0.0, 0.0}

\author{Daniele Malitesta}
\orcid{0000-0003-2228-0333}
\authornote{Corresponding author.}
\affiliation{%
\institution{Université Paris-Saclay CentraleSupélec, Inria}
  \city{Gif-sur-Yvette}
  \country{France}}
\email{daniele.malitesta@centralesupelec.fr}

\author{Emanuele Rossi}
\orcid{0009-0008-5882-4519}
\affiliation{%
\institution{VantAI}
  \city{Barcelona}
  \country{Spain}}
\email{emanuele.rossi1909@gmail.com}

\author{Claudio Pomo}
\orcid{0000-0001-5206-3909}
\affiliation{%
\institution{Politecnico di Bari}
  \city{Bari}
  \country{Italy}}
\email{claudio.pomo@poliba.it}

\author{Tommaso {Di Noia}}
\orcid{0000-0002-0939-5462}
\affiliation{%
\institution{Politecnico di Bari}
  \city{Bari}
  \country{Italy}}
\email{tommaso.dinoia@poliba.it}

\author{Fra\-gkiskos D. Malliaros}
\orcid{0000-0002-8770-3969}
\affiliation{%
\institution{Université Paris-Saclay CentraleSupélec, Inria}
  \city{Gif-sur-Yvette}
  \country{France}}
\email{fragkiskos.malliaros@centralesupelec.fr}

\copyrightyear{2024}
\acmYear{2024}
\setcopyright{rightsretained}
\acmConference[CIKM '24]{Proceedings of the 33rd ACM International Conference on Information and Knowledge Management}{October 21--25, 2024}{Boise, ID, USA}
\acmBooktitle{Proceedings of the 33rd ACM International Conference on Information and Knowledge Management (CIKM '24), October 21--25, 2024, Boise, ID, USA}
\acmDOI{10.1145/3627673.3679898}
\acmISBN{979-8-4007-0436-9/24/10}

\makeatletter
\gdef\@copyrightpermission{
  \begin{minipage}{0.3\columnwidth}
   \href{https://creativecommons.org/licenses/by/4.0/}{\includegraphics[width=0.90\textwidth]{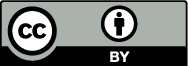}}
  \end{minipage}\hfill
  \begin{minipage}{0.7\columnwidth}
   \href{https://creativecommons.org/licenses/by/4.0/}{This work is licensed under a Creative Commons Attribution International 4.0 License.}
  \end{minipage}
  \vspace{5pt}
}
\makeatother

\renewcommand\footnotetextcopyrightpermission[1]{}

\begin{document}

\title{Do We Really Need to Drop Items with Missing Modalities in Multimodal Recommendation?}

\renewcommand{\shortauthors}{Daniele Malitesta, Emanuele Rossi, Claudio Pomo, Tommaso Di Noia, \& Fragkiskos D. Malliaros}



\begin{abstract}

Generally, items with missing modalities are dropped in multimodal recommendation. However, with this work, we question this procedure, highlighting that it would further damage the pipeline of any multimodal recommender system. First, we show that the lack of (some) modalities is, in fact, a widely-diffused phenomenon in multimodal recommendation. Second, we propose a pipeline that imputes missing multimodal features in recommendation by leveraging traditional imputation strategies in machine learning. Then, given the graph structure of the recommendation data, we also propose three more effective imputation solutions that leverage the item-item co-purchase graph and the multimodal similarities of co-interacted items. Our method can be plugged into any multimodal RSs in the literature working as an untrained pre-processing phase, showing (through extensive experiments) that any data pre-filtering is not only unnecessary but also harmful to the performance.
\end{abstract}

\maketitle

\section{Introduction and motivation}
\label{sec:introduction}

Despite the large success of recommender systems (RSs) algorithms built around the paradigm of collaborative filtering~\cite{DBLP:journals/fthci/EkstrandRK11} (CF), they still face several challenges. Among those, it is acknowledged that RSs struggle when learning from very \textit{sparse} user-item data~\cite{DBLP:conf/aaai/HeM16}, a widely common setting in real-world applications. In such contexts, the literature suggests to enhance the information carried by the user-item data through additional \textit{side information}, such as \textit{multimodal} data (e.g., images, descriptions, audio tracks) accompanying items in several domains (e.g., fashion~\cite{DBLP:conf/cikm/AnelliDNSFMP22, DBLP:conf/kdd/ChenHXGGSLPZZ19,DBLP:conf/ecir/DeldjooNMM22}, music~\cite{DBLP:conf/sigir/ChengSH16,DBLP:conf/recsys/OramasNSS17,DBLP:conf/bigmm/VaswaniAA21}, food~\cite{DBLP:journals/tmm/MinJJ20,DBLP:journals/eswa/LeiHZSZ21,DBLP:journals/tomccap/WangDJJSN21}, and micro-video~\cite{DBLP:conf/mm/WeiWN0HC19,DBLP:journals/tmm/ChenLXZ21,DBLP:journals/tmm/CaiQFX22} recommendation). Multimodal RSs~\cite{DBLP:conf/mm/MalitestaGPN23} leverage high-level features extracted from multimodal items' data to inject/process them into the recommendation pipeline, outperforming CF recommendation~\cite{DBLP:conf/aaai/HeM16, DBLP:conf/mm/WeiWN0HC19, DBLP:conf/mm/WeiWN0C20, DBLP:conf/mm/Zhang00WWW21, DBLP:conf/mm/ZhouS23, DBLP:conf/www/ZhouZLZMWYJ23, DBLP:conf/www/WeiHXZ23}. 

While the multimodal recommendation pipeline hugely relies on the \textit{quality} of multimodal data to extract meaningful items' features~\cite{DBLP:conf/mm/MalitestaGPN23, 10.1145/3662738, DBLP:conf/cvpr/DeldjooNMM21}, in real-world scenarios, such data may be \textit{noisy} or (even worse) \textbf{missing}~\cite{DBLP:journals/corr/abs-2403-19841}. Concretely, if we consider the e-commerce scenario, it is not rare to observe inaccessible URLs of product images and unavailable descriptions/reviews on the product webpage. Even assuming that this phenomenon only applies to real-world recommendation data, a simple analysis of a popular dataset for multimodal recommendation (the Amazon Reviews data~\cite{DBLP:conf/sigir/McAuleyTSH15}) reveals that this is also true in academic research (\Cref{tab:datasets}). 

\begin{table}[!t]
    \caption{Datasets statistics, where we show items with missing visual (V) and textual (T) information, before and after they have been dropped (without and with the apex).}\label{tab:datasets}
    \centering
    \footnotesize
    \begin{tabular}{lrrrrr}
    \toprule
        \multirow{2}{*}{\textbf{Datasets}} & \multirow{2}{*}{\textbf{$|\mathcal{U}|\;/\;|\mathcal{U}'|$}} & \multirow{2}{*}{\textbf{$|\mathcal{I}|\;/\;|\mathcal{I}'|$}} & \multirow{2}{*}{\textbf{$|\mathcal{R}|\;/\;|\mathcal{R}'|$}} & \multicolumn{2}{c}{\textbf{Missing}} \\ \cmidrule{5-6}
        & & & &  V & T \\
        \cmidrule{1-6}
        Office & 4,905 / 4,891 & 2,420 / 1,746 & 53,258 / 35,185 & 0 & 674\\ 
        Music & 5,541 / 5,349 & 3,568 / 2,453 & 64,706 / 51,516 & 2 & 1,114\\
        Beauty & 22,363 / 22,293 & 12,101 / 11,124 & 198,502 / 165,772 & 7 & 977 \\
        \bottomrule
    \end{tabular}
\end{table}

In machine learning, the literature recognizes a whole research line addressing \textit{missing information}~\cite{DBLP:phd/ca/Marlin08, DBLP:journals/jbd/EmmanuelMMSMT21, DBLP:journals/air/LinT20}. The same applies to RSs, where works have focused on missing user-item interactions~\cite{DBLP:journals/siamrev/Strawderman89, 10.1093/biomet/63.3.581, DBLP:conf/kdd/Steck10, DBLP:conf/ijcai/MarlinZRS11,  DBLP:conf/recsys/LimML15, DBLP:conf/recsys/YangCXWBE18, DBLP:conf/nips/WangGZZ18, DBLP:journals/tkde/ZhengWXLW22, DBLP:conf/icml/WangZ0Q19, DBLP:conf/wsdm/SaitoYNSN20, DBLP:conf/iclr/LiZ023} and (in very few cases) users' and items' metadata~\cite{DBLP:conf/cikm/ShiZYZHLM19, DBLP:journals/tkde/LiuCZLN22}. However, even acknowledging the extensive literature on \textbf{missing multimodal information} in deep learning~\cite{DBLP:conf/cvpr/0002R0T022, DBLP:conf/aaai/MaRZTWP21, DBLP:conf/cvpr/LeeTCL23, DBLP:conf/acii/JaquesTSP17, DBLP:journals/taffco/WagnerALK11, DBLP:conf/kdd/ZhangCMZWWZ22, DBLP:conf/sigir/ZengL022, DBLP:journals/tifs/Marin-JimenezCD21}, to date, very limited effort has been devoted to this issue in multimodal recommendation. The only work  is~\cite{DBLP:conf/emnlp/WangNL18}, an \textbf{end-to-end} framework that uses modalities correlations and multimodal dropout during training, and a reconstruction autoencoder for the inference. 

Nevertheless, in most cases, multimodal RSs pre-process the data by dropping items with missing multimodal information. This procedure inevitably leads to losing further data regarding users and user-item interactions in the system (\Cref{tab:datasets}), which is completely \textit{counteractive} to the idea of enhancing the \textit{sparse} user-item recommendation data through multimodal information. Hence, the question is: ``\textbf{Do we really need to drop items with missing modalities in multimodal recommendation?''}. 
The short answer is \textbf{no}, as we empirically demonstrate in this work (\Cref{tab:rq1}), showing that the performance improvement of multimodal RSs (e.g., VBPR~\cite{DBLP:conf/aaai/HeM16}) over their pure CF versions (e.g., BPRMF~\cite{DBLP:conf/uai/RendleFGS09}) can even widen when, as a \textbf{pre-processing}, we impute the items' missing modalities through \textbf{untrained} strategies instead of dropping them.

To this end, in this work, we provide the following contributions. (1) We formalize the understudied problem of \textit{missing modalities} in multimodal recommendation. (2) We propose a pipeline that imputes the items' missing multimodal features acting as an \textit{untrained pre-processing} procedure. (3) Alongside \textit{traditional} machine learning imputations, we propose three novel \textbf{graph-aware} imputation methods that leverage the user-item graph, and take inspiration from \textbf{node features propagation}~\cite{DBLP:conf/log/RossiK0C0B22} and \textbf{diffusion}~\cite{DBLP:conf/nips/KlicperaWG19} in graph learning. (4) Through extensive experiments, we prove that dropping items with missing modalities in multimodal recommendation is \textbf{not only unnecessary but also harmful}; we further support our findings and proposal through an ablation study and hyper-parameter sensitivity analysis to assess the goodness of the proposed graph-aware imputation methods. 

\section{Methodology}
\label{sec:methodology}

In this section, we present our proposed pipeline where missing multimodal features are imputed in an \textit{untrained pre-processing} manner. First, we formalize the missing modalities setting as observed in the collected recommendation datasets. Then, we describe the imputation methods to recover missing modalities, categorized as \textit{traditional} and our proposed \textit{graph-aware} ones.

\subsection{Missing modalities in recommendation}

\textbf{\ul{Preliminaries.}} In a recommendation system, let $\mathcal{U}$ and $\mathcal{I}$ be the sets of users and items, respectively. Then, we indicate with $\mathbf{R} \in \mathbb{R}^{|\mathcal{U}| \times |\mathcal{I}|}$ the user-item interaction matrix, where $\mathbf{R}_{ui} = 1$ if there exists a recorded implicit feedback of user $u \in \mathcal{U}$ over item $i \in \mathcal{I}$, 0 otherwise. As in any latent factor-based recommendation approach~\cite{DBLP:journals/computer/KorenBV09}, we denote with $\mathbf{E}_u \in \mathbb{R}^d$ and $\mathbf{E}_i \in \mathbb{R}^d$ the $d$-dimensional embeddings for user $u \in \mathcal{U}$ and item $i \in \mathcal{I}$. 

In a multimodal recommendation setting, items' representation may be suitably enriched through their multimodal features. Thus, we introduce $\mathcal{M}$ as the set of modalities, and $\mathbf{F} \in \mathbb{R}^{|\mathcal{I}| \times |\mathcal{M}| \times c}$ as the multimodal feature tensor of all items, where $\mathbf{F}_{im} \in \mathbb{R}^{c}$ is the feature of item $i$ accounting for the $m$ modality.

\noindent \textbf{\ul{Missing modalities}.} As already observed in~\Cref{tab:datasets}, the problem of missing modalities in multimodal recommendation implies the unavailability of \textbf{some of (or the whole) multimodal feature vectors} for a specific subset of items. For instance, in a double modalities setting (e.g., visual and textual) we might have that item $i$ is missing either the visual or textual features or both; in every such case, we say item $i$ has some missing modalities, and indicate the missing multimodal features as $\mathbf{F}'_{im} \in \mathbb{R}^c$ for every $m \in \mathcal{M}$ that is missing. Note that the considered setting implies the presence/missingness of the whole feature vector $\mathbf{F}'_{im} \in \mathbb{R}^{c}$.

\subsection{Modalities imputation methods}

\textbf{\ul{Traditional imputation.}} The problem of missing data information in machine learning has been historically discussed and addressed for decades. Following this research line, we adapt three simple and common solutions to impute missing modalities: (i) \textsc{Zeros}, where a vector of all zero entries replaces missing multimodal features; (ii) \textsc{Random}, where a vector of all random entries replaces missing multimodal features; (iii) \textsc{GlobalMean}, where missing multimodal features are replaced by the mean of available multimodal features. 

\noindent \textbf{\ul{Graph-aware imputation.}}
Differently from other data types, recommendation data is a bipartite and undirected graph of user-item interactions. Thus, we leverage the topology of the user-item recommendation graph to propose more enhanced imputing strategies. 

Specifically, we aim to impute missing multimodal features of an item through the existing (i.e., non-missing) multimodal features of the connected items in the \textbf{item-item co-purchase graph} at multiple hops. Indeed, our assumption (supported by several works~\cite{DBLP:conf/mm/Zhang00WWW21,DBLP:conf/mm/LiuYLWTZSM21,DBLP:conf/mir/LiuMSO022, DBLP:conf/mm/ZhouS23}) is that \textbf{co-interacted items are likely to present semantically-similar multimodal features}, aligning with the same principle behind collaborative filtering (users sharing similar preferences tend to interact with the same items). The item-item co-interaction matrix is obtained as $\mathbf{R}^{\mathcal{I}} = \mathbf{R}^\top \mathbf{R}$, where $\mathbf{R}^{\mathcal{I}} \in \mathbb{R}^{|\mathcal{I}| \times |\mathcal{I}|}$ and $\mathbf{R}^{\mathcal{I}}_{ij}$ is the number of users who interacted with both items $i$ and $j$. We sparsify it through top-$k$ sparsification~\cite{DBLP:conf/mm/Zhang00WWW21} obtaining $\overline{\mathbf{R}}^{\mathcal{I}}$ as the item-item sparsified matrix.

Then, inspired by recent works from graph learning~\cite{DBLP:conf/log/RossiK0C0B22}, we propose to frame the imputation of missing modalities in multimodal recommendation as a process involving the \textbf{propagation of nodes multimodal features over the item-item graph}. Specifically, we aim to recover the nodes missing multimodal features through the propagation (at multiple hops) of the existing multimodal features of their neighborhood nodes in the item-item graph.

Let $\mathbf{F}_{im}'$ be the missing multimodal features for item $i$ and modality $m$, and $\mathcal{N}_i = \{j \;|\; \overline{\mathbf{R}}_{ij}^{\mathcal{I}} = 1\}$ be the set of neighborhood nodes of $i$ in the item-item graph. By leveraging feature propagation from the nearest neighborhood nodes in the item-item graph (i.e., one hop), we extend the \textsc{GlobalMean} imputation through the mean of multimodal features from the one-hop neighborhood (\textsc{NeighMean}): $\mathbf{F}_{im}' = \frac{\sum_{j \in \mathcal{N}_i} \mathbf{F}_{jm}}{|\mathcal{N}_i|}, \forall m \in \mathcal{M}$.
Since item $i$ may be linked to items with missing modalities, $\mathbf{F}_{jm}$ would not be known in advance. For this reason, before running the propagation, we initialize all multimodal features of items with missing modalities to the zero vector.

\begin{table*}[!t]
\caption{Performance improvement of VBPR and NGCF-M over BPRMF and NGCF in the \textbf{dropped} and \textbf{imputed} settings.}\label{tab:rq1}
    \centering
    \footnotesize
    \begin{tabular}{lcccccccccccc}
    \toprule
        \multirow{3}{*}{\textbf{Models}} & \multicolumn{6}{c}{\textbf{Dropped}} & \multicolumn{6}{c}{\textbf{Imputed (Ours)}} \\
        \cmidrule(lr){2-7} \cmidrule(lr){8-13}
        & \multicolumn{2}{c}{\textbf{Office}} & \multicolumn{2}{c}{\textbf{Music}} & \multicolumn{2}{c}{\textbf{Beauty}} &  \multicolumn{2}{c}{\textbf{Office}} & \multicolumn{2}{c}{\textbf{Music}} & \multicolumn{2}{c}{\textbf{Beauty}} \\ \cmidrule(lr){2-3} \cmidrule(lr){4-5} \cmidrule(lr){6-7} \cmidrule(lr){8-9} \cmidrule(lr){10-11} \cmidrule(lr){12-13}
        & Recall & nDCG & Recall & nDCG & Recall & nDCG & Recall & nDCG & Recall & nDCG & Recall & nDCG \\
        \cmidrule{1-13}
        BPRMF & 7.74 & 3.65 & 26.50 & 14.24 & 9.81 & 5.00 & 8.51 & 4.55 & 24.59 & 13.69 & 9.43 & 4.85 \\
        VBPR & 9.99 & 4.82 & 25.22 & 13.71 & 11.12 & 5.86 & 10.09 & 5.51 & 26.25 & 14.58 & 10.93 & 5.74 \\ \cmidrule{1-13}
        Improvement (\%) & \textcolor{green}{+29.07\%} & \textcolor{green}{+32.05\%} & \textcolor{red}{-4.83\%} & \textcolor{red}{-3.72\%} & \textcolor{green}{+13.35\%} & \textcolor{green}{+17.20\%} & \textcolor{green}{+18.57\%} & \textcolor{green}{+21.10\%} & \textcolor{green}{+6.75\%}* & \textcolor{green}{+6.50\%}* & \textcolor{green}{+15.91\%}* & \textcolor{green}{+18.35\%}* \\ \cmidrule{1-13}
        NGCF & 9.90 & 4.49 & 24.39 & 12.61 & 8.64 & 4.28 & 8.18 & 3.91 & 24.12 & 13.06 & 8.92 & 4.56 \\
        NGCF-M & 13.10 & 6.36 & 24.12 & 12.73 & 9.68 & 4.94 & 11.65 & 5.85 & 25.18 & 13.64 & 11.00 & 5.76 \\ \cmidrule{1-13}
       Improvement (\%) & \textcolor{green}{+31.32\%} & \textcolor{green}{+41.65\%} & \textcolor{red}{-1.11\%} & \textcolor{green}{+0.95\%} & \textcolor{green}{+12.04\%} & \textcolor{green}{+15.42\%} & \textcolor{green}{+42.42\%}* & \textcolor{green}{+49.62\%}* & \textcolor{green}{+4.39\%}* & \textcolor{green}{+4.44\%}* & \textcolor{green}{+23.32\%}* & \textcolor{green}{+26.32\%}* \\
        \bottomrule
        \multicolumn{13}{l}{\footnotesize * \textit{The performance improvement is higher in our imputed setting than in the dropped one.}}
    \end{tabular}
\end{table*}

As we leverage the item-item graph, we can also consider \textbf{multi-hops} relationships between items. Thus, our second graph-aware method \textbf{iteratively} performs feature propagation (\textsc{MultiHop}): $\mathbf{F}_{im}'^{(t)} = \frac{\sum_{j \in \mathcal{N}_i} \mathbf{F}_{jm}^{(t-1)}}{\sqrt{|\mathcal{N}_i|}\sqrt{|\mathcal{N}_j|}}, \forall m \in \mathcal{M}, \forall t \in \{1, \dots, T\}$.
Here, $\mathbf{F}_{jm}^{(0)}$ is again initialized to the zero vector for all items with missing modalities. Then, we set $\mathbf{F}_{jm}^{(t - 1)} = \mathbf{F}_{jm}^{(0)}$ for items with existing modalities as, at each new iteration $t \in \{1, \dots, T\}$, we want to re-initialize the multimodal features to their original values~\cite{DBLP:conf/log/RossiK0C0B22}.

Above, we perform the symmetric Laplacian normalization of the item-item co-interaction matrix to smooth the noisy item-item connections we initially built. 
While this operation only accounts for the 1-hop neighborhoods, the recent literature has demonstrated that \textbf{graph diffusion at multiple hops}~\cite{DBLP:conf/nips/KlicperaWG19} may improve graph learning; the idea is to leverage an enhanced representation of the interaction matrix, as personalized PageRank~\cite{ilprints422}.

Hence, we propose a third graph-aware imputation that modifies the \textsc{MultiHop} strategy by incorporating \textbf{personalized PageRank} as a normalization of the item-item co-interactions matrix. Let $\mathbf{B} \in \mathbb{R}^{|\mathcal{I}| \times |\mathcal{I}|}$ be a matrix such that $\mathbf{B}_{ij} = 1 - (1 - \alpha) / (\sqrt{|\mathcal{N}_i|}\sqrt{|\mathcal{N}_j|})$ if $i = j$, $\mathbf{B}_{ij} = - (1 - \alpha) / (\sqrt{|\mathcal{N}_i|}\sqrt{|\mathcal{N}_j|})$ otherwise. Then, we have (\textsc{PersPageRank}): $\mathbf{F}_{im}'^{(t)} =  \sum_{j \in \mathcal{N}_i} (\alpha \mathbf{B}^{-1})_{ij} \mathbf{F}_{jm}^{(t-1)}, \forall m \in \mathcal{M}, \forall t \in \{1, \dots, T\}$,
where $\alpha \in [0, 1]$ is the teleport probability of going from an item $i$ to the 1-hop connected item $j$ in the item-item graph. 

\section{Experiments and results}
In this section, we report on the experiments and discuss the results. First, we describe the datasets, and baselines, and provide reproducibility details. Then, we answer four research questions: \textbf{RQ1.)} What is the performance improvement of multimodal recommender systems over their pure collaborative filtering versions in both the dropped and imputed dataset settings? \textbf{RQ2.)} What is the impact of missing modalities imputation on recent multimodal RSs? \textbf{RQ3.)} How does each imputation strategy contribute to the performance? \textbf{RQ4.)} What is the effect of top-$k$ sparsification and the number of propagation hops on the performance?

\subsection{Experimental settings}

\noindent \textbf{\ul{Datasets.}} We select three categories of the popular Amazon Reviews dataset~\cite{DBLP:conf/sigir/McAuleyTSH15}, namely: Office Products (\textbf{Products}), Digital Music (\textbf{Music}), and \textbf{Beauty} (\Cref{tab:datasets}). For each of them, we consider two distinct versions: (i) \textbf{dropped}, where we remove items with at least one missing modality; (ii) \textbf{imputed}, where we impute the items' missing multimodal features. From the available items' multimodal data (i.e., product images and textual descriptions), we extract~\cite{DBLP:conf/www/AttimonelliDMPG24} visual~\cite{DBLP:conf/cvpr/HeZRS16} and textual embeddings~\cite{DBLP:conf/emnlp/ReimersG19}.

\noindent\textbf{\ul{Baselines.}} We select CF and multimodal models: BPRMF~\cite{DBLP:conf/uai/RendleFGS09}, NGCF~\cite{DBLP:conf/sigir/Wang0WFC19}, LightGCN~\cite{DBLP:conf/sigir/0001DWLZ020}, SGL~\cite{DBLP:conf/sigir/WuWF0CLX21}, VBPR~\cite{DBLP:conf/aaai/HeM16}, NGCF-M, FREEDOM~\cite{DBLP:conf/mm/ZhouS23}, and BM3~\cite{DBLP:conf/www/ZhouZLZMWYJ23}. We introduce NGCF-M as the multimodal version of NGCF where multimodal features are injected in the same manner as VBPR for BPRMF. For the imputation, we use three \textit{classical machine learning} methods (i.e., \textsc{Zeros}, \textsc{Random}, \textsc{GlobalMean}), and propose three novel \textit{graph-aware} approaches (i.e., \textsc{NeighMean}, \textsc{MultiHop}, and \textsc{PersPageRank}). 

\noindent \textbf{\ul{Reproducibility.}} We split the datasets into train and test sets (80\%/20\%) and run the experiments with~\cite{10.1145/3662738}. We explore the learning rate in \{0.0001, 0.0005, 0.001, 0.005, 0.01\} and the regularization term in [1e-5, 1e-2], leaving the other parameters to the best values, and fix the batch size at 1024 and the epochs at 200. Then, we explore the top-$k$ sparsification in [10, 20, \dots, 100] and the propagation hops $T$ in [1, 2, \dots, 20]. We use the Recall@20 and the nDCG@20 as test recommendation metrics and retain 10\% of the train set as validation, with Recall@20 as a validation metric. The code is at:~\url{https://github.com/sisinflab/Graph-Missing-Modalities}.   
\subsection{Results and discussion}

\noindent \textbf{\ul{Multimodal improvement over CF (RQ1).}}
\Cref{tab:rq1} displays the performance improvement of VBPR and NGCF-M over their CF versions (i.e., BPRMF and NGCF) in the \textbf{dropped} and \textbf{imputed} recommendation settings. Note that we decide to consider VBPR and NGCF-M for the current analysis as, differently from the other multimodal baselines we will use later, they differ from their respective CF versions for the \textbf{sole} injection of items' multimodal features. Thus, we can precisely assess the impact of multimodal features imputation and avoid the risk that other factors (e.g., denoising procedures as in FREEDOM~\cite{DBLP:conf/mm/ZhouS23}) could inadvertently influence the results. Moreover, since the dropped and imputed versions of each dataset may largely differ (refer again to~\Cref{tab:datasets}), we are mainly interested in the \textbf{performance improvement} between CF and multimodal RSs; we decide to report also the performance for all models/datasets pairs only for the sake of completeness. 

In the vast majority of cases, we observe that the multimodal version outperforms the CF one. Nevertheless, while such improvement may sometimes not occur in the dropped setting (see Music), the imputed setting \textbf{consistently} shows an improvement from CF to multimodal RSs. Additionally, the imputed performance improvement is always quite \textbf{higher in magnitude} than the dropped one (refer to the asteriscs in the table). We assume that the imputed setting permits (i) retaining important information in the recommendation data (i.e., users, items, and interactions) and (ii) recovering missing multimodal features in a high-quality manner.

\begin{table}[!t]
\caption{Performance in the imputed setting. Boldface and underline are the best and second-best values.}\label{tab:rq2}
    \centering
    \footnotesize
    \begin{tabular}{lcccccc}
    \toprule
        \multirow{2}{*}{\textbf{Models}} &  \multicolumn{2}{c}{\textbf{Office}} & \multicolumn{2}{c}{\textbf{Music}} & \multicolumn{2}{c}{\textbf{Beauty}} \\ \cmidrule(lr){2-3} \cmidrule(lr){4-5} \cmidrule(lr){6-7}
        & Recall & nDCG & Recall & nDCG & Recall & nDCG \\ \cmidrule{1-7}
        BPRMF & 8.51 & 4.55 & 24.59 & 13.69 & 9.43 & 4.85 \\
        NGCF & 8.18 & 3.91 & 24.12 & 13.06 & 8.92 & 4.56 \\
        LightGCN & 11.55 & 6.12 & 26.61 & 14.49 & 11.61 & 6.02 \\
        SGL & 10.28 & 5.81 & \textbf{27.59} & \textbf{15.86} & \underline{12.20} & \underline{6.72} \\
        VBPR & 10.09 & 5.51 & 26.25 & 14.58 & 10.93 & 5.74 \\
        NGCF-M & \underline{11.65} & 5.85 & 25.18 & 13.64 & 11.00 & 5.76 \\
        FREEDOM & \textbf{12.79} & \textbf{6.76} & \underline{27.18} & \underline{14.91} & \textbf{13.28} & \textbf{7.00} \\
        BM3 & 11.28 & \underline{6.15} & 25.11 & 13.55 & 11.49 & 6.04 \\
        \bottomrule
    \end{tabular}
\end{table}

\noindent \textbf{\ul{Impact on recent multimodal RSs (RQ2).}}
We test if the imputation of missing modalities in \textbf{recent} multimodal RSs can lead to performance trends that are aligned with the related literature. To this aim, we extend the set of RSs with LightGCN, SGL (pure CF), and FREEDOM, BM3 (multimodal). \Cref{tab:rq2} displays the recommendation results in the \textbf{imputed} setting.  Overall, we notice two trends aligned with the literature. On the one hand, more recent approaches (e.g., LightGCN and SGL on the CF side, FREEDOM and BM3 on the multimodal side) can often provide higher performance. On the other hand, the adoption of refined approaches leveraging multimodality (e.g., FREEDOM and BM3) can frequently lead to improved performance over pure CF (e.g., see Office and Beauty).

\noindent \textbf{\ul{Ablation study (RQ3).}}
\Cref{tab:rq3} shows the performance of FREEDOM and BM3 on Music and Beauty, highlighting the various imputation methods. Overall, we note that our proposed \textbf{graph-aware} imputation methods (i.e., \textsc{NeighMean}, \textsc{MultiHop}, \textsc{PersPageRank}) always \textbf{outperform} traditional machine learning imputation (i.e., \textsc{Zeros}, \textsc{Random}, and \textsc{GlobalMean}); as expected, we see \textsc{MultiHop} and \textsc{PersPageRank} steadily settle as the superior solutions. Thus, we can effectively leverage the \textbf{graph structure} of the user-item interaction data to adopt more tailored and refined graph-aware imputation methods that exploit item-item co-interactions and semantic similarities, especially at multiple distance hops.

\begin{table}[!t]
\caption{Ablation study with varying imputation methods.}\label{tab:rq3}
    \centering
    \footnotesize
    \begin{tabular}{llcccc}
    \toprule
        \multirow{2}{*}{\textbf{Models}} & \multirow{2}{*}{\textbf{Imputation}} & \multicolumn{2}{c}{\textbf{Music}} & \multicolumn{2}{c}{\textbf{Beauty}} \\ \cmidrule(lr){3-4} \cmidrule(lr){5-6} 
        & & Recall & nDCG & Recall & nDCG \\ \cmidrule{1-6}
        \multirow{6}{*}{FREEDOM} & +\textsc{Zeros} & 25.60 & 13.78 & 10.99 & 5.76 \\
        & +\textsc{Random} & 26.81 & 14.46 & 13.00 & 6.73 \\
        & +\textsc{GlobalMean} & 26.62 & 14.61 & 13.02 & 6.80 \\
        & +\textsc{NeighMean} & 26.17 & 14.35 & 13.23 & 6.88 \\
        & +\textsc{MultiHop} & \underline{27.09} & \textbf{14.91} & \underline{13.26} & \underline{6.96} \\
        & +\textsc{PersPageRank} & \textbf{27.18} & \underline{14.88} & \textbf{13.28} & \textbf{7.00} \\
        \cmidrule{1-6}
        \multirow{6}{*}{BM3} & +\textsc{Zeros} & 24.64 & 13.00 & 11.31 & 5.96 \\
        & +\textsc{Random} & 24.29 & 12.93 & 11.35 & 5.88  \\
        & +\textsc{GlobalMean} & 24.27 & 12.88 & 11.35 & 5.91 \\
        & +\textsc{NeighMean} & 24.46 & 12.95 & \underline{11.48} & \underline{6.01} \\
        & +\textsc{MultiHop} & \underline{25.00} & \textbf{13.55} & \textbf{11.49} & \textbf{6.04} \\
        & +\textsc{PersPageRank} & \textbf{25.11} & \underline{13.44} & 11.35 & 5.96 \\
        \bottomrule
    \end{tabular}
\end{table}

\begin{figure}[!t]
    \centering
    \includegraphics[width=\columnwidth]{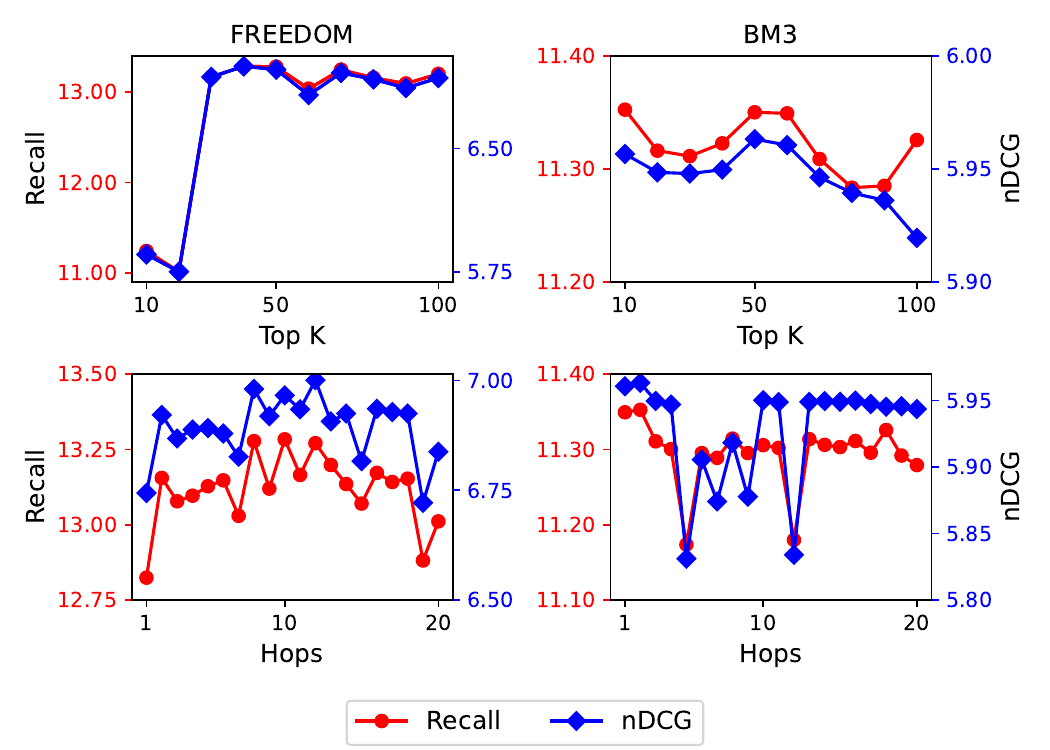}
    \caption{Impact of top-$k$ sparsification (top) and propagation hops (bottom) on the Beauty dataset for \textsc{PersPageRank}.}
    \label{fig:rq4}
\end{figure}

\noindent \textbf{\ul{Hyper-parameter sensitivity (RQ4).}}
Finally, we assess the impact of top-$k$ sparsification and propagation layers for the best \textbf{graph-aware} imputation strategy, \textsc{PersPageRank}. \Cref{fig:rq4} shows the performance changes when varying the top-$k$ sparsification (top) and the propagation layers (bottom) on Beauty. Regarding top-$k$ sparsification, we obtain the best results on FREEDOM for $k \geq 30$, while results on BM3 are better around $k = 50$. Regarding the propagation layers, the outcomes suggest that on FREEDOM we reach the highest metric values around $T = 10$, while on BM3 the trend is almost stable for $T \geq 10$. Conclusively, \textbf{high} top-$k$ \textbf{sparsifications} and many propagation hops can improve the results on both FREEDOM and BM3, as they remove \textbf{noisy} item-item \textbf{interactions} and leverage \textbf{multiple hops} in the item-item graph.
\section{Conclusion and future work}
\label{sec:conclusion}

Unlike the common practice of dropping items with missing modalities in multimodal recommendation, in this work, we proposed an untrained pre-processing pipeline to impute missing multimodal features through traditional methods and novel graph-aware strategies. First, we showed that imputing can preserve (or even improve) the performance gap between multimodal and pure collaborative recommender systems. Then, we demonstrated the superior efficacy of our proposed graph-aware imputation methods. We plan to further extend the experimental setting to additional datasets and techniques, as well as design new imputation solutions to integrate end-to-end in the recommendation pipeline~\cite{DBLP:conf/emnlp/WangNL18}.

\begin{acks}
Daniele Malitesta and Fragkiskos D. Malliaros acknowledge the support by ANR (French National Research Agency) under the JCJC project GraphIA (ANR-20-CE23-0009-01). Claudio Pomo and Tommaso Di Noia acknowledge the partial support by CT\_FINCONS\_III, OVS Fashion Retail Reloaded, LUTECH DIGITALE 4.0, Secure Safe Apulia, IDENTITA, REACH-XY. The authors acknowledge Inria and the CINECA award under the ISCRA initiative for the availability of high-performance computing resources.
\end{acks}

\balance
\bibliographystyle{ACM-Reference-Format}
\bibliography{bibliography}


\begin{thebibliography}{62}


\ifx \showCODEN    \undefined \def \showCODEN     #1{\unskip}     \fi
\ifx \showDOI      \undefined \def \showDOI       #1{#1}\fi
\ifx \showISBNx    \undefined \def \showISBNx     #1{\unskip}     \fi
\ifx \showISBNxiii \undefined \def \showISBNxiii  #1{\unskip}     \fi
\ifx \showISSN     \undefined \def \showISSN      #1{\unskip}     \fi
\ifx \showLCCN     \undefined \def \showLCCN      #1{\unskip}     \fi
\ifx \shownote     \undefined \def \shownote      #1{#1}          \fi
\ifx \showarticletitle \undefined \def \showarticletitle #1{#1}   \fi
\ifx \showURL      \undefined \def \showURL       {\relax}        \fi
\providecommand\bibfield[2]{#2}
\providecommand\bibinfo[2]{#2}
\providecommand\natexlab[1]{#1}
\providecommand\showeprint[2][]{arXiv:#2}

\bibitem[Anelli et~al\mbox{.}(2022)]%
        {DBLP:conf/cikm/AnelliDNSFMP22}
\bibfield{author}{\bibinfo{person}{Vito~Walter Anelli}, \bibinfo{person}{Yashar Deldjoo}, \bibinfo{person}{Tommaso~Di Noia}, \bibinfo{person}{Eugenio~Di Sciascio}, \bibinfo{person}{Antonio Ferrara}, \bibinfo{person}{Daniele Malitesta}, {and} \bibinfo{person}{Claudio Pomo}.} \bibinfo{year}{2022}\natexlab{}.
\newblock \showarticletitle{Reshaping Graph Recommendation with Edge Graph Collaborative Filtering and Customer Reviews}. In \bibinfo{booktitle}{\emph{DL4SR@CIKM}} \emph{(\bibinfo{series}{{CEUR} Workshop Proceedings}, Vol.~\bibinfo{volume}{3317})}. \bibinfo{publisher}{CEUR-WS.org}.
\newblock


\bibitem[Attimonelli et~al\mbox{.}(2024)]%
        {DBLP:conf/www/AttimonelliDMPG24}
\bibfield{author}{\bibinfo{person}{Matteo Attimonelli}, \bibinfo{person}{Danilo Danese}, \bibinfo{person}{Daniele Malitesta}, \bibinfo{person}{Claudio Pomo}, \bibinfo{person}{Giuseppe Gassi}, {and} \bibinfo{person}{Tommaso~Di Noia}.} \bibinfo{year}{2024}\natexlab{}.
\newblock \showarticletitle{Ducho 2.0: Towards a More Up-to-Date Unified Framework for the Extraction of Multimodal Features in Recommendation}. In \bibinfo{booktitle}{\emph{{WWW} (Companion Volume)}}. \bibinfo{publisher}{{ACM}}, \bibinfo{pages}{1075--1078}.
\newblock


\bibitem[Cai et~al\mbox{.}(2022)]%
        {DBLP:journals/tmm/CaiQFX22}
\bibfield{author}{\bibinfo{person}{Desheng Cai}, \bibinfo{person}{Shengsheng Qian}, \bibinfo{person}{Quan Fang}, {and} \bibinfo{person}{Changsheng Xu}.} \bibinfo{year}{2022}\natexlab{}.
\newblock \showarticletitle{Heterogeneous Hierarchical Feature Aggregation Network for Personalized Micro-Video Recommendation}.
\newblock \bibinfo{journal}{\emph{{IEEE} Trans. Multim.}}  \bibinfo{volume}{24} (\bibinfo{year}{2022}), \bibinfo{pages}{805--818}.
\newblock


\bibitem[Chen et~al\mbox{.}(2019)]%
        {DBLP:conf/kdd/ChenHXGGSLPZZ19}
\bibfield{author}{\bibinfo{person}{Wen Chen}, \bibinfo{person}{Pipei Huang}, \bibinfo{person}{Jiaming Xu}, \bibinfo{person}{Xin Guo}, \bibinfo{person}{Cheng Guo}, \bibinfo{person}{Fei Sun}, \bibinfo{person}{Chao Li}, \bibinfo{person}{Andreas Pfadler}, \bibinfo{person}{Huan Zhao}, {and} \bibinfo{person}{Binqiang Zhao}.} \bibinfo{year}{2019}\natexlab{}.
\newblock \showarticletitle{{POG:} Personalized Outfit Generation for Fashion Recommendation at Alibaba iFashion}. In \bibinfo{booktitle}{\emph{{KDD}}}. \bibinfo{publisher}{{ACM}}.
\newblock


\bibitem[Chen et~al\mbox{.}(2021)]%
        {DBLP:journals/tmm/ChenLXZ21}
\bibfield{author}{\bibinfo{person}{Xusong Chen}, \bibinfo{person}{Dong Liu}, \bibinfo{person}{Zhiwei Xiong}, {and} \bibinfo{person}{Zheng{-}Jun Zha}.} \bibinfo{year}{2021}\natexlab{}.
\newblock \showarticletitle{Learning and Fusing Multiple User Interest Representations for Micro-Video and Movie Recommendations}.
\newblock \bibinfo{journal}{\emph{{IEEE} Trans. Multim.}}  \bibinfo{volume}{23} (\bibinfo{year}{2021}), \bibinfo{pages}{484--496}.
\newblock


\bibitem[Cheng et~al\mbox{.}(2016)]%
        {DBLP:conf/sigir/ChengSH16}
\bibfield{author}{\bibinfo{person}{Zhiyong Cheng}, \bibinfo{person}{Jialie Shen}, {and} \bibinfo{person}{Steven C.~H. Hoi}.} \bibinfo{year}{2016}\natexlab{}.
\newblock \showarticletitle{On Effective Personalized Music Retrieval by Exploring Online User Behaviors}. In \bibinfo{booktitle}{\emph{{SIGIR}}}. \bibinfo{publisher}{{ACM}}, \bibinfo{pages}{125--134}.
\newblock


\bibitem[Deldjoo et~al\mbox{.}(2021)]%
        {DBLP:conf/cvpr/DeldjooNMM21}
\bibfield{author}{\bibinfo{person}{Yashar Deldjoo}, \bibinfo{person}{Tommaso~Di Noia}, \bibinfo{person}{Daniele Malitesta}, {and} \bibinfo{person}{Felice~Antonio Merra}.} \bibinfo{year}{2021}\natexlab{}.
\newblock \showarticletitle{A Study on the Relative Importance of Convolutional Neural Networks in Visually-Aware Recommender Systems}. In \bibinfo{booktitle}{\emph{{CVPR} Workshops}}. \bibinfo{publisher}{Computer Vision Foundation / {IEEE}}, \bibinfo{pages}{3961--3967}.
\newblock


\bibitem[Deldjoo et~al\mbox{.}(2022)]%
        {DBLP:conf/ecir/DeldjooNMM22}
\bibfield{author}{\bibinfo{person}{Yashar Deldjoo}, \bibinfo{person}{Tommaso~Di Noia}, \bibinfo{person}{Daniele Malitesta}, {and} \bibinfo{person}{Felice~Antonio Merra}.} \bibinfo{year}{2022}\natexlab{}.
\newblock \showarticletitle{Leveraging Content-Style Item Representation for Visual Recommendation}. In \bibinfo{booktitle}{\emph{{ECIR} {(2)}}} \emph{(\bibinfo{series}{Lecture Notes in Computer Science}, Vol.~\bibinfo{volume}{13186})}. \bibinfo{publisher}{Springer}, \bibinfo{pages}{84--92}.
\newblock


\bibitem[Ekstrand et~al\mbox{.}(2011)]%
        {DBLP:journals/fthci/EkstrandRK11}
\bibfield{author}{\bibinfo{person}{Michael~D. Ekstrand}, \bibinfo{person}{John Riedl}, {and} \bibinfo{person}{Joseph~A. Konstan}.} \bibinfo{year}{2011}\natexlab{}.
\newblock \showarticletitle{Collaborative Filtering Recommender Systems}.
\newblock \bibinfo{journal}{\emph{Found. Trends Hum. Comput. Interact.}} \bibinfo{volume}{4}, \bibinfo{number}{2} (\bibinfo{year}{2011}), \bibinfo{pages}{175--243}.
\newblock


\bibitem[Emmanuel et~al\mbox{.}(2021)]%
        {DBLP:journals/jbd/EmmanuelMMSMT21}
\bibfield{author}{\bibinfo{person}{Tlamelo Emmanuel}, \bibinfo{person}{Thabiso~M. Maupong}, \bibinfo{person}{Dimane Mpoeleng}, \bibinfo{person}{Thabo Semong}, \bibinfo{person}{Banyatsang Mphago}, {and} \bibinfo{person}{Oteng Tabona}.} \bibinfo{year}{2021}\natexlab{}.
\newblock \showarticletitle{A survey on missing data in machine learning}.
\newblock \bibinfo{journal}{\emph{J. Big Data}} \bibinfo{volume}{8}, \bibinfo{number}{1} (\bibinfo{year}{2021}), \bibinfo{pages}{140}.
\newblock


\bibitem[He et~al\mbox{.}(2016)]%
        {DBLP:conf/cvpr/HeZRS16}
\bibfield{author}{\bibinfo{person}{Kaiming He}, \bibinfo{person}{Xiangyu Zhang}, \bibinfo{person}{Shaoqing Ren}, {and} \bibinfo{person}{Jian Sun}.} \bibinfo{year}{2016}\natexlab{}.
\newblock \showarticletitle{Deep Residual Learning for Image Recognition}. In \bibinfo{booktitle}{\emph{{CVPR}}}. \bibinfo{publisher}{{IEEE} Computer Society}, \bibinfo{pages}{770--778}.
\newblock


\bibitem[He and McAuley(2016)]%
        {DBLP:conf/aaai/HeM16}
\bibfield{author}{\bibinfo{person}{Ruining He} {and} \bibinfo{person}{Julian~J. McAuley}.} \bibinfo{year}{2016}\natexlab{}.
\newblock \showarticletitle{{VBPR:} Visual Bayesian Personalized Ranking from Implicit Feedback}. In \bibinfo{booktitle}{\emph{{AAAI}}}. \bibinfo{publisher}{{AAAI} Press}, \bibinfo{pages}{144--150}.
\newblock


\bibitem[He et~al\mbox{.}(2020)]%
        {DBLP:conf/sigir/0001DWLZ020}
\bibfield{author}{\bibinfo{person}{Xiangnan He}, \bibinfo{person}{Kuan Deng}, \bibinfo{person}{Xiang Wang}, \bibinfo{person}{Yan Li}, \bibinfo{person}{Yong{-}Dong Zhang}, {and} \bibinfo{person}{Meng Wang}.} \bibinfo{year}{2020}\natexlab{}.
\newblock \showarticletitle{LightGCN: Simplifying and Powering Graph Convolution Network for Recommendation}. In \bibinfo{booktitle}{\emph{{SIGIR}}}. \bibinfo{publisher}{{ACM}}, \bibinfo{pages}{639--648}.
\newblock


\bibitem[Jaques et~al\mbox{.}(2017)]%
        {DBLP:conf/acii/JaquesTSP17}
\bibfield{author}{\bibinfo{person}{Natasha Jaques}, \bibinfo{person}{Sara Taylor}, \bibinfo{person}{Akane Sano}, {and} \bibinfo{person}{Rosalind~W. Picard}.} \bibinfo{year}{2017}\natexlab{}.
\newblock \showarticletitle{Multimodal autoencoder: {A} deep learning approach to filling in missing sensor data and enabling better mood prediction}. In \bibinfo{booktitle}{\emph{{ACII}}}. \bibinfo{publisher}{{IEEE} Computer Society}, \bibinfo{pages}{202--208}.
\newblock


\bibitem[Klicpera et~al\mbox{.}(2019)]%
        {DBLP:conf/nips/KlicperaWG19}
\bibfield{author}{\bibinfo{person}{Johannes Klicpera}, \bibinfo{person}{Stefan Wei{\ss}enberger}, {and} \bibinfo{person}{Stephan G{\"{u}}nnemann}.} \bibinfo{year}{2019}\natexlab{}.
\newblock \showarticletitle{Diffusion Improves Graph Learning}. In \bibinfo{booktitle}{\emph{NeurIPS}}. \bibinfo{pages}{13333--13345}.
\newblock


\bibitem[Koren et~al\mbox{.}(2009)]%
        {DBLP:journals/computer/KorenBV09}
\bibfield{author}{\bibinfo{person}{Yehuda Koren}, \bibinfo{person}{Robert~M. Bell}, {and} \bibinfo{person}{Chris Volinsky}.} \bibinfo{year}{2009}\natexlab{}.
\newblock \showarticletitle{Matrix Factorization Techniques for Recommender Systems}.
\newblock \bibinfo{journal}{\emph{Computer}} \bibinfo{volume}{42}, \bibinfo{number}{8} (\bibinfo{year}{2009}), \bibinfo{pages}{30--37}.
\newblock


\bibitem[Lee et~al\mbox{.}(2023)]%
        {DBLP:conf/cvpr/LeeTCL23}
\bibfield{author}{\bibinfo{person}{Yi{-}Lun Lee}, \bibinfo{person}{Yi{-}Hsuan Tsai}, \bibinfo{person}{Wei{-}Chen Chiu}, {and} \bibinfo{person}{Chen{-}Yu Lee}.} \bibinfo{year}{2023}\natexlab{}.
\newblock \showarticletitle{Multimodal Prompting with Missing Modalities for Visual Recognition}. In \bibinfo{booktitle}{\emph{{CVPR}}}. \bibinfo{publisher}{{IEEE}}, \bibinfo{pages}{14943--14952}.
\newblock


\bibitem[Lei et~al\mbox{.}(2021)]%
        {DBLP:journals/eswa/LeiHZSZ21}
\bibfield{author}{\bibinfo{person}{Zhenfeng Lei}, \bibinfo{person}{Anwar~Ul Haq}, \bibinfo{person}{Adnan Zeb}, \bibinfo{person}{Md Suzauddola}, {and} \bibinfo{person}{Defu Zhang}.} \bibinfo{year}{2021}\natexlab{}.
\newblock \showarticletitle{Is the suggested food your desired?: Multi-modal recipe recommendation with demand-based knowledge graph}.
\newblock \bibinfo{journal}{\emph{Expert Syst. Appl.}}  \bibinfo{volume}{186} (\bibinfo{year}{2021}), \bibinfo{pages}{115708}.
\newblock


\bibitem[Li et~al\mbox{.}(2023)]%
        {DBLP:conf/iclr/LiZ023}
\bibfield{author}{\bibinfo{person}{Haoxuan Li}, \bibinfo{person}{Chunyuan Zheng}, {and} \bibinfo{person}{Peng Wu}.} \bibinfo{year}{2023}\natexlab{}.
\newblock \showarticletitle{StableDR: Stabilized Doubly Robust Learning for Recommendation on Data Missing Not at Random}. In \bibinfo{booktitle}{\emph{{ICLR}}}. \bibinfo{publisher}{OpenReview.net}.
\newblock


\bibitem[Lim et~al\mbox{.}(2015)]%
        {DBLP:conf/recsys/LimML15}
\bibfield{author}{\bibinfo{person}{Daryl Lim}, \bibinfo{person}{Julian~J. McAuley}, {and} \bibinfo{person}{Gert R.~G. Lanckriet}.} \bibinfo{year}{2015}\natexlab{}.
\newblock \showarticletitle{Top-N Recommendation with Missing Implicit Feedback}. In \bibinfo{booktitle}{\emph{RecSys}}. \bibinfo{publisher}{{ACM}}, \bibinfo{pages}{309--312}.
\newblock


\bibitem[Lin and Tsai(2020)]%
        {DBLP:journals/air/LinT20}
\bibfield{author}{\bibinfo{person}{Wei{-}Chao Lin} {and} \bibinfo{person}{Chih{-}Fong Tsai}.} \bibinfo{year}{2020}\natexlab{}.
\newblock \showarticletitle{Missing value imputation: a review and analysis of the literature {(2006-2017)}}.
\newblock \bibinfo{journal}{\emph{Artif. Intell. Rev.}} \bibinfo{volume}{53}, \bibinfo{number}{2} (\bibinfo{year}{2020}), \bibinfo{pages}{1487--1509}.
\newblock


\bibitem[Liu et~al\mbox{.}(2022a)]%
        {DBLP:journals/tkde/LiuCZLN22}
\bibfield{author}{\bibinfo{person}{Fan Liu}, \bibinfo{person}{Zhiyong Cheng}, \bibinfo{person}{Lei Zhu}, \bibinfo{person}{Chenghao Liu}, {and} \bibinfo{person}{Liqiang Nie}.} \bibinfo{year}{2022}\natexlab{a}.
\newblock \showarticletitle{An Attribute-Aware Attentive {GCN} Model for Attribute Missing in Recommendation}.
\newblock \bibinfo{journal}{\emph{{IEEE} Trans. Knowl. Data Eng.}} \bibinfo{volume}{34}, \bibinfo{number}{9} (\bibinfo{year}{2022}), \bibinfo{pages}{4077--4088}.
\newblock


\bibitem[Liu et~al\mbox{.}(2021)]%
        {DBLP:conf/mm/LiuYLWTZSM21}
\bibfield{author}{\bibinfo{person}{Yong Liu}, \bibinfo{person}{Susen Yang}, \bibinfo{person}{Chenyi Lei}, \bibinfo{person}{Guoxin Wang}, \bibinfo{person}{Haihong Tang}, \bibinfo{person}{Juyong Zhang}, \bibinfo{person}{Aixin Sun}, {and} \bibinfo{person}{Chunyan Miao}.} \bibinfo{year}{2021}\natexlab{}.
\newblock \showarticletitle{Pre-training Graph Transformer with Multimodal Side Information for Recommendation}. In \bibinfo{booktitle}{\emph{{ACM} Multimedia}}. \bibinfo{publisher}{{ACM}}, \bibinfo{pages}{2853--2861}.
\newblock


\bibitem[Liu et~al\mbox{.}(2022b)]%
        {DBLP:conf/mir/LiuMSO022}
\bibfield{author}{\bibinfo{person}{Zhuang Liu}, \bibinfo{person}{Yunpu Ma}, \bibinfo{person}{Matthias Schubert}, \bibinfo{person}{Yuanxin Ouyang}, {and} \bibinfo{person}{Zhang Xiong}.} \bibinfo{year}{2022}\natexlab{b}.
\newblock \showarticletitle{Multi-Modal Contrastive Pre-training for Recommendation}. In \bibinfo{booktitle}{\emph{{ICMR}}}. \bibinfo{publisher}{{ACM}}, \bibinfo{pages}{99--108}.
\newblock


\bibitem[Ma et~al\mbox{.}(2022)]%
        {DBLP:conf/cvpr/0002R0T022}
\bibfield{author}{\bibinfo{person}{Mengmeng Ma}, \bibinfo{person}{Jian Ren}, \bibinfo{person}{Long Zhao}, \bibinfo{person}{Davide Testuggine}, {and} \bibinfo{person}{Xi Peng}.} \bibinfo{year}{2022}\natexlab{}.
\newblock \showarticletitle{Are Multimodal Transformers Robust to Missing Modality?}. In \bibinfo{booktitle}{\emph{{CVPR}}}. \bibinfo{publisher}{{IEEE}}, \bibinfo{pages}{18156--18165}.
\newblock


\bibitem[Ma et~al\mbox{.}(2021)]%
        {DBLP:conf/aaai/MaRZTWP21}
\bibfield{author}{\bibinfo{person}{Mengmeng Ma}, \bibinfo{person}{Jian Ren}, \bibinfo{person}{Long Zhao}, \bibinfo{person}{Sergey Tulyakov}, \bibinfo{person}{Cathy Wu}, {and} \bibinfo{person}{Xi Peng}.} \bibinfo{year}{2021}\natexlab{}.
\newblock \showarticletitle{{SMIL:} Multimodal Learning with Severely Missing Modality}. In \bibinfo{booktitle}{\emph{{AAAI}}}. \bibinfo{publisher}{{AAAI} Press}, \bibinfo{pages}{2302--2310}.
\newblock


\bibitem[Malitesta et~al\mbox{.}(2024a)]%
        {10.1145/3662738}
\bibfield{author}{\bibinfo{person}{Daniele Malitesta}, \bibinfo{person}{Giandomenico Cornacchia}, \bibinfo{person}{Claudio Pomo}, \bibinfo{person}{Felice~Antonio Merra}, \bibinfo{person}{Tommaso Di~Noia}, {and} \bibinfo{person}{Eugenio Di~Sciascio}.} \bibinfo{year}{2024}\natexlab{a}.
\newblock \showarticletitle{Formalizing Multimedia Recommendation through Multimodal Deep Learning}.
\newblock \bibinfo{journal}{\emph{ACM Trans. Recomm. Syst.}} (\bibinfo{date}{apr} \bibinfo{year}{2024}).
\newblock
\urldef\tempurl%
\url{https://doi.org/10.1145/3662738}
\showDOI{\tempurl}
\newblock
\shownote{Just Accepted}.


\bibitem[Malitesta et~al\mbox{.}(2023)]%
        {DBLP:conf/mm/MalitestaGPN23}
\bibfield{author}{\bibinfo{person}{Daniele Malitesta}, \bibinfo{person}{Giuseppe Gassi}, \bibinfo{person}{Claudio Pomo}, {and} \bibinfo{person}{Tommaso~Di Noia}.} \bibinfo{year}{2023}\natexlab{}.
\newblock \showarticletitle{Ducho: {A} Unified Framework for the Extraction of Multimodal Features in Recommendation}. In \bibinfo{booktitle}{\emph{{ACM} Multimedia}}. \bibinfo{publisher}{{ACM}}, \bibinfo{pages}{9668--9671}.
\newblock


\bibitem[Malitesta et~al\mbox{.}(2024b)]%
        {DBLP:journals/corr/abs-2403-19841}
\bibfield{author}{\bibinfo{person}{Daniele Malitesta}, \bibinfo{person}{Emanuele Rossi}, \bibinfo{person}{Claudio Pomo}, \bibinfo{person}{Fragkiskos~D. Malliaros}, {and} \bibinfo{person}{Tommaso~Di Noia}.} \bibinfo{year}{2024}\natexlab{b}.
\newblock \showarticletitle{Dealing with Missing Modalities in Multimodal Recommendation: a Feature Propagation-based Approach}.
\newblock \bibinfo{journal}{\emph{CoRR}}  \bibinfo{volume}{abs/2403.19841} (\bibinfo{year}{2024}).
\newblock


\bibitem[Mar{\'{\i}}n{-}Jim{\'{e}}nez et~al\mbox{.}(2021)]%
        {DBLP:journals/tifs/Marin-JimenezCD21}
\bibfield{author}{\bibinfo{person}{Manuel~J. Mar{\'{\i}}n{-}Jim{\'{e}}nez}, \bibinfo{person}{Francisco~M. Castro}, \bibinfo{person}{Rub{\'{e}}n Delgado{-}Esca{\~{n}}o}, \bibinfo{person}{Vicky Kalogeiton}, {and} \bibinfo{person}{Nicol{\'{a}}s Guil}.} \bibinfo{year}{2021}\natexlab{}.
\newblock \showarticletitle{UGaitNet: Multimodal Gait Recognition With Missing Input Modalities}.
\newblock \bibinfo{journal}{\emph{{IEEE} Trans. Inf. Forensics Secur.}}  \bibinfo{volume}{16} (\bibinfo{year}{2021}), \bibinfo{pages}{5452--5462}.
\newblock


\bibitem[Marlin(2008)]%
        {DBLP:phd/ca/Marlin08}
\bibfield{author}{\bibinfo{person}{Benjamin~M. Marlin}.} \bibinfo{year}{2008}\natexlab{}.
\newblock \emph{\bibinfo{title}{Missing Data Problems in Machine Learning}}.
\newblock \bibinfo{thesistype}{Ph.\,D. Dissertation}. \bibinfo{school}{University of Toronto, Canada}.
\newblock


\bibitem[Marlin et~al\mbox{.}(2011)]%
        {DBLP:conf/ijcai/MarlinZRS11}
\bibfield{author}{\bibinfo{person}{Benjamin~M. Marlin}, \bibinfo{person}{Richard~S. Zemel}, \bibinfo{person}{Sam~T. Roweis}, {and} \bibinfo{person}{Malcolm Slaney}.} \bibinfo{year}{2011}\natexlab{}.
\newblock \showarticletitle{Recommender Systems, Missing Data and Statistical Model Estimation}. In \bibinfo{booktitle}{\emph{{IJCAI}}}. \bibinfo{publisher}{{IJCAI/AAAI}}, \bibinfo{pages}{2686--2691}.
\newblock


\bibitem[McAuley et~al\mbox{.}(2015)]%
        {DBLP:conf/sigir/McAuleyTSH15}
\bibfield{author}{\bibinfo{person}{Julian~J. McAuley}, \bibinfo{person}{Christopher Targett}, \bibinfo{person}{Qinfeng Shi}, {and} \bibinfo{person}{Anton van~den Hengel}.} \bibinfo{year}{2015}\natexlab{}.
\newblock \showarticletitle{Image-Based Recommendations on Styles and Substitutes}. In \bibinfo{booktitle}{\emph{{SIGIR}}}. \bibinfo{publisher}{{ACM}}, \bibinfo{pages}{43--52}.
\newblock


\bibitem[Min et~al\mbox{.}(2020)]%
        {DBLP:journals/tmm/MinJJ20}
\bibfield{author}{\bibinfo{person}{Weiqing Min}, \bibinfo{person}{Shuqiang Jiang}, {and} \bibinfo{person}{Ramesh~C. Jain}.} \bibinfo{year}{2020}\natexlab{}.
\newblock \showarticletitle{Food Recommendation: Framework, Existing Solutions, and Challenges}.
\newblock \bibinfo{journal}{\emph{{IEEE} Trans. Multim.}} \bibinfo{volume}{22}, \bibinfo{number}{10} (\bibinfo{year}{2020}), \bibinfo{pages}{2659--2671}.
\newblock


\bibitem[Oramas et~al\mbox{.}(2017)]%
        {DBLP:conf/recsys/OramasNSS17}
\bibfield{author}{\bibinfo{person}{Sergio Oramas}, \bibinfo{person}{Oriol Nieto}, \bibinfo{person}{Mohamed Sordo}, {and} \bibinfo{person}{Xavier Serra}.} \bibinfo{year}{2017}\natexlab{}.
\newblock \showarticletitle{A Deep Multimodal Approach for Cold-start Music Recommendation}. In \bibinfo{booktitle}{\emph{DLRS@RecSys}}. \bibinfo{publisher}{{ACM}}, \bibinfo{pages}{32--37}.
\newblock


\bibitem[Page et~al\mbox{.}(1999)]%
        {ilprints422}
\bibfield{author}{\bibinfo{person}{Lawrence Page}, \bibinfo{person}{Sergey Brin}, \bibinfo{person}{Rajeev Motwani}, {and} \bibinfo{person}{Terry Winograd}.} \bibinfo{year}{1999}\natexlab{}.
\newblock \bibinfo{booktitle}{\emph{The PageRank Citation Ranking: Bringing Order to the Web.}}
\newblock \bibinfo{type}{Technical Report} 1999-66. \bibinfo{institution}{Stanford InfoLab}.
\newblock


\bibitem[Reimers and Gurevych(2019)]%
        {DBLP:conf/emnlp/ReimersG19}
\bibfield{author}{\bibinfo{person}{Nils Reimers} {and} \bibinfo{person}{Iryna Gurevych}.} \bibinfo{year}{2019}\natexlab{}.
\newblock \showarticletitle{Sentence-BERT: Sentence Embeddings using Siamese BERT-Networks}. In \bibinfo{booktitle}{\emph{{EMNLP/IJCNLP} {(1)}}}. \bibinfo{publisher}{Association for Computational Linguistics}, \bibinfo{pages}{3980--3990}.
\newblock


\bibitem[Rendle et~al\mbox{.}(2009)]%
        {DBLP:conf/uai/RendleFGS09}
\bibfield{author}{\bibinfo{person}{Steffen Rendle}, \bibinfo{person}{Christoph Freudenthaler}, \bibinfo{person}{Zeno Gantner}, {and} \bibinfo{person}{Lars Schmidt{-}Thieme}.} \bibinfo{year}{2009}\natexlab{}.
\newblock \showarticletitle{{BPR:} Bayesian Personalized Ranking from Implicit Feedback}. In \bibinfo{booktitle}{\emph{{UAI}}}.
\newblock


\bibitem[Rossi et~al\mbox{.}(2022)]%
        {DBLP:conf/log/RossiK0C0B22}
\bibfield{author}{\bibinfo{person}{Emanuele Rossi}, \bibinfo{person}{Henry Kenlay}, \bibinfo{person}{Maria~I. Gorinova}, \bibinfo{person}{Benjamin~Paul Chamberlain}, \bibinfo{person}{Xiaowen Dong}, {and} \bibinfo{person}{Michael~M. Bronstein}.} \bibinfo{year}{2022}\natexlab{}.
\newblock \showarticletitle{On the Unreasonable Effectiveness of Feature Propagation in Learning on Graphs With Missing Node Features}. In \bibinfo{booktitle}{\emph{LoG}} \emph{(\bibinfo{series}{Proceedings of Machine Learning Research}, Vol.~\bibinfo{volume}{198})}. \bibinfo{publisher}{{PMLR}}, \bibinfo{pages}{11}.
\newblock


\bibitem[RUBIN(1976)]%
        {10.1093/biomet/63.3.581}
\bibfield{author}{\bibinfo{person}{DONALD~B. RUBIN}.} \bibinfo{year}{1976}\natexlab{}.
\newblock \showarticletitle{{Inference and missing data}}.
\newblock \bibinfo{journal}{\emph{Biometrika}} \bibinfo{volume}{63}, \bibinfo{number}{3} (\bibinfo{date}{12} \bibinfo{year}{1976}), \bibinfo{pages}{581--592}.
\newblock
\showISSN{0006-3444}
\urldef\tempurl%
\url{https://doi.org/10.1093/biomet/63.3.581}
\showDOI{\tempurl}
\showeprint{https://academic.oup.com/biomet/article-pdf/63/3/581/756166/63-3-581.pdf}


\bibitem[Saito et~al\mbox{.}(2020)]%
        {DBLP:conf/wsdm/SaitoYNSN20}
\bibfield{author}{\bibinfo{person}{Yuta Saito}, \bibinfo{person}{Suguru Yaginuma}, \bibinfo{person}{Yuta Nishino}, \bibinfo{person}{Hayato Sakata}, {and} \bibinfo{person}{Kazuhide Nakata}.} \bibinfo{year}{2020}\natexlab{}.
\newblock \showarticletitle{Unbiased Recommender Learning from Missing-Not-At-Random Implicit Feedback}. In \bibinfo{booktitle}{\emph{{WSDM}}}. \bibinfo{publisher}{{ACM}}, \bibinfo{pages}{501--509}.
\newblock


\bibitem[Shi et~al\mbox{.}(2019)]%
        {DBLP:conf/cikm/ShiZYZHLM19}
\bibfield{author}{\bibinfo{person}{Shaoyun Shi}, \bibinfo{person}{Min Zhang}, \bibinfo{person}{Xinxing Yu}, \bibinfo{person}{Yongfeng Zhang}, \bibinfo{person}{Bin Hao}, \bibinfo{person}{Yiqun Liu}, {and} \bibinfo{person}{Shaoping Ma}.} \bibinfo{year}{2019}\natexlab{}.
\newblock \showarticletitle{Adaptive Feature Sampling for Recommendation with Missing Content Feature Values}. In \bibinfo{booktitle}{\emph{{CIKM}}}. \bibinfo{publisher}{{ACM}}, \bibinfo{pages}{1451--1460}.
\newblock


\bibitem[Steck(2010)]%
        {DBLP:conf/kdd/Steck10}
\bibfield{author}{\bibinfo{person}{Harald Steck}.} \bibinfo{year}{2010}\natexlab{}.
\newblock \showarticletitle{Training and testing of recommender systems on data missing not at random}. In \bibinfo{booktitle}{\emph{{KDD}}}. \bibinfo{publisher}{{ACM}}, \bibinfo{pages}{713--722}.
\newblock


\bibitem[Strawderman(1989)]%
        {DBLP:journals/siamrev/Strawderman89}
\bibfield{author}{\bibinfo{person}{William~E. Strawderman}.} \bibinfo{year}{1989}\natexlab{}.
\newblock \showarticletitle{Statistical Analysis with Missing Data (Roderick J. A. Little and Donald B. Rubin)}.
\newblock \bibinfo{journal}{\emph{{SIAM} Rev.}} \bibinfo{volume}{31}, \bibinfo{number}{2} (\bibinfo{year}{1989}), \bibinfo{pages}{348--349}.
\newblock


\bibitem[Vaswani et~al\mbox{.}(2021)]%
        {DBLP:conf/bigmm/VaswaniAA21}
\bibfield{author}{\bibinfo{person}{Kunal Vaswani}, \bibinfo{person}{Yudhik Agrawal}, {and} \bibinfo{person}{Vinoo Alluri}.} \bibinfo{year}{2021}\natexlab{}.
\newblock \showarticletitle{Multimodal Fusion Based Attentive Networks for Sequential Music Recommendation}. In \bibinfo{booktitle}{\emph{BigMM}}. \bibinfo{publisher}{{IEEE}}, \bibinfo{pages}{25--32}.
\newblock


\bibitem[Wagner et~al\mbox{.}(2011)]%
        {DBLP:journals/taffco/WagnerALK11}
\bibfield{author}{\bibinfo{person}{Johannes Wagner}, \bibinfo{person}{Elisabeth Andr{\'{e}}}, \bibinfo{person}{Florian Lingenfelser}, {and} \bibinfo{person}{Jonghwa Kim}.} \bibinfo{year}{2011}\natexlab{}.
\newblock \showarticletitle{Exploring Fusion Methods for Multimodal Emotion Recognition with Missing Data}.
\newblock \bibinfo{journal}{\emph{{IEEE} Trans. Affect. Comput.}} \bibinfo{volume}{2}, \bibinfo{number}{4} (\bibinfo{year}{2011}), \bibinfo{pages}{206--218}.
\newblock


\bibitem[Wang et~al\mbox{.}(2018b)]%
        {DBLP:conf/emnlp/WangNL18}
\bibfield{author}{\bibinfo{person}{Cheng Wang}, \bibinfo{person}{Mathias Niepert}, {and} \bibinfo{person}{Hui Li}.} \bibinfo{year}{2018}\natexlab{b}.
\newblock \showarticletitle{{LRMM:} Learning to Recommend with Missing Modalities}. In \bibinfo{booktitle}{\emph{{EMNLP}}}. \bibinfo{publisher}{Association for Computational Linguistics}, \bibinfo{pages}{3360--3370}.
\newblock


\bibitem[Wang et~al\mbox{.}(2018a)]%
        {DBLP:conf/nips/WangGZZ18}
\bibfield{author}{\bibinfo{person}{Menghan Wang}, \bibinfo{person}{Mingming Gong}, \bibinfo{person}{Xiaolin Zheng}, {and} \bibinfo{person}{Kun Zhang}.} \bibinfo{year}{2018}\natexlab{a}.
\newblock \showarticletitle{Modeling Dynamic Missingness of Implicit Feedback for Recommendation}. In \bibinfo{booktitle}{\emph{NeurIPS}}. \bibinfo{pages}{6670--6679}.
\newblock


\bibitem[Wang et~al\mbox{.}(2021)]%
        {DBLP:journals/tomccap/WangDJJSN21}
\bibfield{author}{\bibinfo{person}{Wenjie Wang}, \bibinfo{person}{Ling{-}Yu Duan}, \bibinfo{person}{Hao Jiang}, \bibinfo{person}{Peiguang Jing}, \bibinfo{person}{Xuemeng Song}, {and} \bibinfo{person}{Liqiang Nie}.} \bibinfo{year}{2021}\natexlab{}.
\newblock \showarticletitle{Market2Dish: Health-aware Food Recommendation}.
\newblock \bibinfo{journal}{\emph{{ACM} Trans. Multim. Comput. Commun. Appl.}} \bibinfo{volume}{17}, \bibinfo{number}{1} (\bibinfo{year}{2021}), \bibinfo{pages}{33:1--33:19}.
\newblock


\bibitem[Wang et~al\mbox{.}(2019a)]%
        {DBLP:conf/sigir/Wang0WFC19}
\bibfield{author}{\bibinfo{person}{Xiang Wang}, \bibinfo{person}{Xiangnan He}, \bibinfo{person}{Meng Wang}, \bibinfo{person}{Fuli Feng}, {and} \bibinfo{person}{Tat{-}Seng Chua}.} \bibinfo{year}{2019}\natexlab{a}.
\newblock \showarticletitle{Neural Graph Collaborative Filtering}. In \bibinfo{booktitle}{\emph{{SIGIR}}}. \bibinfo{publisher}{{ACM}}, \bibinfo{pages}{165--174}.
\newblock


\bibitem[Wang et~al\mbox{.}(2019b)]%
        {DBLP:conf/icml/WangZ0Q19}
\bibfield{author}{\bibinfo{person}{Xiaojie Wang}, \bibinfo{person}{Rui Zhang}, \bibinfo{person}{Yu Sun}, {and} \bibinfo{person}{Jianzhong Qi}.} \bibinfo{year}{2019}\natexlab{b}.
\newblock \showarticletitle{Doubly Robust Joint Learning for Recommendation on Data Missing Not at Random}. In \bibinfo{booktitle}{\emph{{ICML}}} \emph{(\bibinfo{series}{Proceedings of Machine Learning Research}, Vol.~\bibinfo{volume}{97})}. \bibinfo{publisher}{{PMLR}}, \bibinfo{pages}{6638--6647}.
\newblock


\bibitem[Wei et~al\mbox{.}(2023)]%
        {DBLP:conf/www/WeiHXZ23}
\bibfield{author}{\bibinfo{person}{Wei Wei}, \bibinfo{person}{Chao Huang}, \bibinfo{person}{Lianghao Xia}, {and} \bibinfo{person}{Chuxu Zhang}.} \bibinfo{year}{2023}\natexlab{}.
\newblock \showarticletitle{Multi-Modal Self-Supervised Learning for Recommendation}. In \bibinfo{booktitle}{\emph{{WWW}}}. \bibinfo{publisher}{{ACM}}, \bibinfo{pages}{790--800}.
\newblock


\bibitem[Wei et~al\mbox{.}(2020)]%
        {DBLP:conf/mm/WeiWN0C20}
\bibfield{author}{\bibinfo{person}{Yinwei Wei}, \bibinfo{person}{Xiang Wang}, \bibinfo{person}{Liqiang Nie}, \bibinfo{person}{Xiangnan He}, {and} \bibinfo{person}{Tat{-}Seng Chua}.} \bibinfo{year}{2020}\natexlab{}.
\newblock \showarticletitle{Graph-Refined Convolutional Network for Multimedia Recommendation with Implicit Feedback}. In \bibinfo{booktitle}{\emph{{ACM} Multimedia}}. \bibinfo{publisher}{{ACM}}, \bibinfo{pages}{3541--3549}.
\newblock


\bibitem[Wei et~al\mbox{.}(2019)]%
        {DBLP:conf/mm/WeiWN0HC19}
\bibfield{author}{\bibinfo{person}{Yinwei Wei}, \bibinfo{person}{Xiang Wang}, \bibinfo{person}{Liqiang Nie}, \bibinfo{person}{Xiangnan He}, \bibinfo{person}{Richang Hong}, {and} \bibinfo{person}{Tat{-}Seng Chua}.} \bibinfo{year}{2019}\natexlab{}.
\newblock \showarticletitle{{MMGCN:} Multi-modal Graph Convolution Network for Personalized Recommendation of Micro-video}. In \bibinfo{booktitle}{\emph{{ACM} Multimedia}}. \bibinfo{publisher}{{ACM}}, \bibinfo{pages}{1437--1445}.
\newblock


\bibitem[Wu et~al\mbox{.}(2021)]%
        {DBLP:conf/sigir/WuWF0CLX21}
\bibfield{author}{\bibinfo{person}{Jiancan Wu}, \bibinfo{person}{Xiang Wang}, \bibinfo{person}{Fuli Feng}, \bibinfo{person}{Xiangnan He}, \bibinfo{person}{Liang Chen}, \bibinfo{person}{Jianxun Lian}, {and} \bibinfo{person}{Xing Xie}.} \bibinfo{year}{2021}\natexlab{}.
\newblock \showarticletitle{Self-supervised Graph Learning for Recommendation}. In \bibinfo{booktitle}{\emph{{SIGIR}}}. \bibinfo{publisher}{{ACM}}, \bibinfo{pages}{726--735}.
\newblock


\bibitem[Yang et~al\mbox{.}(2018)]%
        {DBLP:conf/recsys/YangCXWBE18}
\bibfield{author}{\bibinfo{person}{Longqi Yang}, \bibinfo{person}{Yin Cui}, \bibinfo{person}{Yuan Xuan}, \bibinfo{person}{Chenyang Wang}, \bibinfo{person}{Serge~J. Belongie}, {and} \bibinfo{person}{Deborah Estrin}.} \bibinfo{year}{2018}\natexlab{}.
\newblock \showarticletitle{Unbiased offline recommender evaluation for missing-not-at-random implicit feedback}. In \bibinfo{booktitle}{\emph{RecSys}}. \bibinfo{publisher}{{ACM}}, \bibinfo{pages}{279--287}.
\newblock


\bibitem[Zeng et~al\mbox{.}(2022)]%
        {DBLP:conf/sigir/ZengL022}
\bibfield{author}{\bibinfo{person}{Jiandian Zeng}, \bibinfo{person}{Tianyi Liu}, {and} \bibinfo{person}{Jiantao Zhou}.} \bibinfo{year}{2022}\natexlab{}.
\newblock \showarticletitle{Tag-assisted Multimodal Sentiment Analysis under Uncertain Missing Modalities}. In \bibinfo{booktitle}{\emph{{SIGIR}}}. \bibinfo{publisher}{{ACM}}, \bibinfo{pages}{1545--1554}.
\newblock


\bibitem[Zhang et~al\mbox{.}(2022)]%
        {DBLP:conf/kdd/ZhangCMZWWZ22}
\bibfield{author}{\bibinfo{person}{Chaohe Zhang}, \bibinfo{person}{Xu Chu}, \bibinfo{person}{Liantao Ma}, \bibinfo{person}{Yinghao Zhu}, \bibinfo{person}{Yasha Wang}, \bibinfo{person}{Jiangtao Wang}, {and} \bibinfo{person}{Junfeng Zhao}.} \bibinfo{year}{2022}\natexlab{}.
\newblock \showarticletitle{M3Care: Learning with Missing Modalities in Multimodal Healthcare Data}. In \bibinfo{booktitle}{\emph{{KDD}}}. \bibinfo{publisher}{{ACM}}, \bibinfo{pages}{2418--2428}.
\newblock


\bibitem[Zhang et~al\mbox{.}(2021)]%
        {DBLP:conf/mm/Zhang00WWW21}
\bibfield{author}{\bibinfo{person}{Jinghao Zhang}, \bibinfo{person}{Yanqiao Zhu}, \bibinfo{person}{Qiang Liu}, \bibinfo{person}{Shu Wu}, \bibinfo{person}{Shuhui Wang}, {and} \bibinfo{person}{Liang Wang}.} \bibinfo{year}{2021}\natexlab{}.
\newblock \showarticletitle{Mining Latent Structures for Multimedia Recommendation}. In \bibinfo{booktitle}{\emph{{ACM} Multimedia}}. \bibinfo{publisher}{{ACM}}, \bibinfo{pages}{3872--3880}.
\newblock


\bibitem[Zheng et~al\mbox{.}(2022)]%
        {DBLP:journals/tkde/ZhengWXLW22}
\bibfield{author}{\bibinfo{person}{Xiaolin Zheng}, \bibinfo{person}{Menghan Wang}, \bibinfo{person}{Renjun Xu}, \bibinfo{person}{Jianmeng Li}, {and} \bibinfo{person}{Yan Wang}.} \bibinfo{year}{2022}\natexlab{}.
\newblock \showarticletitle{Modeling Dynamic Missingness of Implicit Feedback for Sequential Recommendation}.
\newblock \bibinfo{journal}{\emph{{IEEE} Trans. Knowl. Data Eng.}} \bibinfo{volume}{34}, \bibinfo{number}{1} (\bibinfo{year}{2022}), \bibinfo{pages}{405--418}.
\newblock


\bibitem[Zhou and Shen(2023)]%
        {DBLP:conf/mm/ZhouS23}
\bibfield{author}{\bibinfo{person}{Xin Zhou} {and} \bibinfo{person}{Zhiqi Shen}.} \bibinfo{year}{2023}\natexlab{}.
\newblock \showarticletitle{A Tale of Two Graphs: Freezing and Denoising Graph Structures for Multimodal Recommendation}. In \bibinfo{booktitle}{\emph{{ACM} Multimedia}}. \bibinfo{publisher}{{ACM}}, \bibinfo{pages}{935--943}.
\newblock


\bibitem[Zhou et~al\mbox{.}(2023)]%
        {DBLP:conf/www/ZhouZLZMWYJ23}
\bibfield{author}{\bibinfo{person}{Xin Zhou}, \bibinfo{person}{Hongyu Zhou}, \bibinfo{person}{Yong Liu}, \bibinfo{person}{Zhiwei Zeng}, \bibinfo{person}{Chunyan Miao}, \bibinfo{person}{Pengwei Wang}, \bibinfo{person}{Yuan You}, {and} \bibinfo{person}{Feijun Jiang}.} \bibinfo{year}{2023}\natexlab{}.
\newblock \showarticletitle{Bootstrap Latent Representations for Multi-modal Recommendation}. In \bibinfo{booktitle}{\emph{{WWW}}}. \bibinfo{publisher}{{ACM}}, \bibinfo{pages}{845--854}.
\newblock


\end{thebibliography}

\end{document}